\documentstyle[floats,twocolumn,aps,psfig,prl]{revtex}

\begin{document}
\draft

%***********    This is for two columns *******************************
\twocolumn[\hsize\textwidth\columnwidth\hsize\csname @twocolumnfalse\endcsname


\title{Stability of antiferromagnetism at high magnetic fields in Mn$_3$Si}

\author{C. Pfleiderer$^1$, J. B\oe uf$^1$, H. v. L\"ohneysen$^{1,2}$}

\address{$^1$Physikalisches Institut, Universit\"at Karlsruhe,
D-76128 Karlsruhe, Germany\\
$^2$Forschungszentrum Karlsruhe, Institut f\"ur Festk\"orperphysik, D-76021 
Karlsruhe, Germany
}

\date{\today}
\maketitle

\begin{abstract}
We report low temperature measurements of the specific heat, resistivity and 
magnetisation of the itinerant antiferromagnet Mn$_3$Si.  The unprecedented stability of the magnetic state to high magnetic field up to 14\,T inferred from the invariance of these bulk properties is incompatible with itinerant magnetism expected of a conventional Fermi liquid. 
\end{abstract}

% body of paper here
% ***********    This is for two columns *******************************
\vskip2pc]

%\subsection{Introduction (a)}

Weakly magnetic transition metal compounds with small ordered moments and low transition temperatures are usually well described as a Fermi liquid with weakly exchange split Fermi surface, where special topological features of the Fermi surface are ignored \cite{Mor85}. It is in contrast still an open issue if magnetic metals with large ordered moments and small ordering temperature, that are more akin to local moment insulators, may also be described on the basis of Fermi liquid theory. Insight into this question may be sought in the dominant energy scales controling the magnetic ordering temperature $T_N$ of an itinerant antiferromagnet. In a Fermi liquid low ordering temperatures result from strong coupling of the magnetisation at ordering wave-vector $\bf Q$ with heavily damped spin fluctuations \cite{Mor85}. On the other hand, in the isotropic Heisenberg model low ordering temperatures result from weak exchange coupling of localized magnetic moments. The energy scale and nature of the magnetic ordering compared to the mode-mode or local moment coupling, respectively, are hence fundamental to magnetism in solids.

The polarization of a magnetic state in a magnetic field allows to probe 
the relevant coupling strengths. In fact, a very large number of transition metal compounds are known in which ordering temperatures of order 20\,K are related to a high sensitivity to magnetic field of a few T regardless of whether they are better described in terms of Fermi liquid theory or the Heisenberg model \cite{Alk95,Str77,MMT}. In this paper we report an experimental investigation in which we employ the effect of high magnetic field ($\mu_B H \approx k_B T_N$) on the bulk magnetisation $M$, specific heat $C$ and 
electrical resistivity $\rho$ of a cubic metallic antiferromagnet, Mn$_3$Si, to study the strength of the coupling of the uniform mode (${\bf q}=0$) with the ordering wave-vector ${\bf Q}$. The observed complete absence of a field dependence is surprising, because it either suggests an isotropic Fermi liquid in which the mode-mode coupling is absent or an isotropic Heisenberg magnet in which the moment coupling is exceptionally strong, despite low $T_N$. From either point of view the conceptual simplicity of Mn$_3$Si challenges the present day understanding of metallic antiferromagnetism.

Recent years have witnessed an increased interest in the magnetic and metallic properties of Heusler transition metal compounds \cite{Kub83,Fuj95,Nis97,Sle00} to which Mn$_3$Si belongs. Ternary members of this class of materials display a wide variety of different behaviours such as disorder induced non-Fermi liquid \cite{Nis97,Sle00}, half-metallic magnetic \cite{Fuj95} and semiconducting ground states \cite{Kub83}. High ferromagnetic ordering temperatures of several hundred K are also frequent.

Mn$_3$Si crystalizes as Mn$_I$Mn$_{II,2}$Si, DO3 ordered Fe$_3$Al, space group Fm3m \cite{Kal74,Bab75,Tom75}. The crystal structure may be described in terms of four fcc sublattices located at (0, 0, 0) for Mn$_I$, (1/4, 1/4, 1/4) and (3/4, 3/4, 3/4) for the Mn$_{II}$ and (1/2, 1/2, 1/2) for Si. The Mn$_I$ sites are hence surrounded by 8 Mn$_{II}$ nearest neighbours (nn) and have an $O_h$ point symmetry. The Mn$_{II}$ are on the other hand surrounded by 4 Mn$_I$ nn and four Si nn, respectively. The structural symmetries encountered in Mn$_3$Si are extremely simple, so that studies of polycrystalline samples already supply key information.

Mn$_3$Si is metallic and develops incommensurate antiferromagnetic order below $T_N\approx 25$\,K along ${\bf Q}$ = 0.425 ${\bf a^*}_{111}$, where ${\bf a^*}_{111}$ denotes the reciprocal lattice vector along $(111)$ \cite{Bab75,Tom75}. The magnetic order is best described as a sequence of ferromagnetic planes Mn$_{II}$ - Mn$_I$ - Mn$_{II}$, separated by a Si plane, all equidistant at $d = 3.304$\,{\AA}. The ordered moments $\mu_I = 1.7 \mu_B$ and $\mu_{II} = 0.19 \mu_B$ are oriented in-plane, so that the sequence of planes form a short wavelength double helix. The ordering wavevector ${\bf Q}$ is weakly $T$ dependent without hint for a further change of magnetic structure \cite{Tom75}. The $T$ dependence of the Mn$_I$ moment follows a Brillouin function ($S=1/2$) for $T<T_N$, i. e., the transition is second order, the magnetic state essentially isotropic and a sizeable ordered moment forms already closely below $T_N$ \cite{Tom75}. The key features of the magnetic structure are that (i) the order is essentially antiferromagnetic, (ii) there is no magnetic site frustration, (iii) there is no appreciable magnetic anisotropy, and (iv) the amplitude of the moments varies strongly by nearly an order of magnitude along ${\bf Q}$. 

Inelastic neutron scattering in Mn$_3$Si shows excitation spectra that are unusual in a number of ways. For $T>T_N$ large amplitude, large coherence-length antiferromagnetic spin fluctuations are observed \cite{Tom87,Yam95}. These
fluctuations exhibit a $\bf q$-dependence that is nearly three orders of magnitude stronger between ${\bf q}=0$ and ${\bf Q}$ than the conventional Lorentzian dependence expected of an isotropic Fermi liquid. For $T<T_N$ very slow antiferromagnetic spin waves with velocity $c_{l}\approx 37$\,meV{\AA}\,$\approx 85$\,km/s around ${\bf Q}$ below 6\,meV contrast fast antiferromagnetic spin waves near the zone boundary above 6\,meV, where $c_{h} \gg c_{l}$ \cite{Tom87}. We note, however, that neither of these excitations qualify properly as propagating spin waves, due to their extreme broadening $\delta\omega/\omega \approx 1$ which contrasts the spin wave broadening conventionally observed in itinerant magnets $\delta\omega/\omega \approx 0.1 - 0.001$ \cite{Col89}. 

\begin{figure}[t]
\centerline{\psfig{file=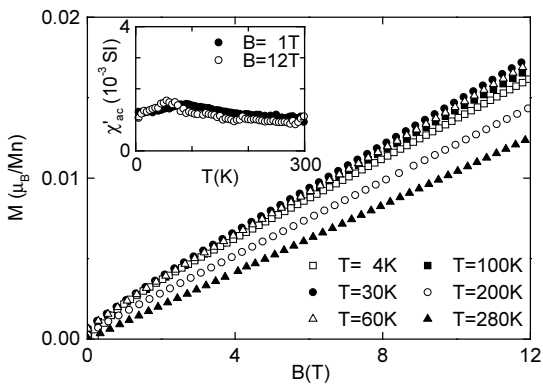,clip=,width=7.5cm}}
\caption{Isothermal magnetisation of Mn$_3$Si as function of magnetic field. The inset displays the $T$ dependence of the a.c. susceptibility at $B=$1\,T and 12\,T. The average magnetic moment at 12\,T is nearly two orders of magnitude smaller than the ordered magnetic moment. The linear field dependence underscores the apparent lack of a coupling between the uniform magnetic field and the ordering wave vector ${\bf Q}$.}
\end{figure}

For our studies polycrystalline samples were synthesised from stoichiometric amounts of high-purity starting materials (5N Mn; 6N Si) by Ar arc-melting. Samples were melted several times to promote homogeneity and subsequently annealed at 700$^{\circ}$\,C for 7\,days. No measurable weight loss was observed. $\Theta- 2\Theta$ powder diffraction suggested that the samples were single phase where the lattice constant $a = 5.722\,${\AA} agreed with the published value. Careful scans across polished faces of the polycrystals using microprobe analysis eventually revealed $< 1 \%$ of a second phase of Mn$_5$Si$_3$. This Si-rich phase may be related to a peritectic decomposition of Mn$_3$Si above 850\,$^{\circ}$\,C. Neither the polycrystalline nature of the samples nor tiny contributions from the second phase of Mn$_5$Si$_3$, for which antiferromagnetic order is very sensitive to magnetic field \cite{Alk95}, alter the key observations and conclusions reported here.

Shown in Fig. 1 is the magnetization $M$ as function of magnetic field $B$ 
as measured in a conventional vibrating sample magnetometer. Up to 12\,T, the highest field studied, only a tiny portion of the ordered moment of less than $3 \%$ is recovered. The magnetic response is linear in $B$ and characteristic of a simple Pauli paramagnetic state over the entire $B$ and $T$ range investigated. The associated susceptibility as measured by a low frequency technique is low even in the immediate vicinity of $T_N$ (inset of Fig.1). This contrasts the $T$ dependence of the ordered Mn$_I$ moment for $T<T_N$ which reaches $\sim$\,70\,$\%$ already a few K below $T_N$, i.\,e., a large increase of $M$ is anticipated for a field induced change of moment orientation just below $T_N$ where the ordering energy is low versus the field energy.

\begin{figure}[t]
\centerline{\psfig{file=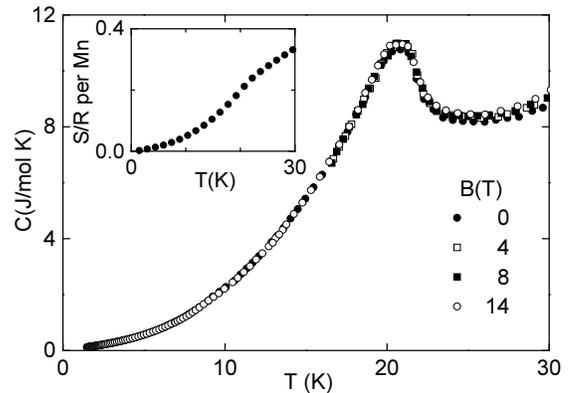,clip=,width=7.5cm}}
\caption{Specific heat as function of temperature at finite field up to 14\,T.
The inset displays the entropy as calculated from the measured specific heat. No magnetic field dependence is observed apart from a very weak increase of order 3$\%$ above $T_N$. The entropy is siginificantly exceeds that of typical weak itinerant magnets.}
\end{figure}

The specific heat $C(T)$ is dominated by a pronounced anomaly just below $T_N$ (Fig. 2). For $T \to 0$ we observe $C(T) \approx  \gamma T + \beta T^3$, where
$\gamma \approx$\,69 mJ/mol K$^2$ and $\beta \approx 1.5 \cdot 10^{-3}$\,J/molK$^4$ is consistent with the zero-field behaviour reported before \cite{Kal74}. The rounding of the anomaly in $C(T)$  may be due to the polycrystalline nature of the samples studied here. Up to a field of 14\,T we do not observe any variation of $C(T)$ within the resolution of the experiment, apart from a slight increase above $T_N$. Any anisotropy would broaden the $C(T)$ considerably since $T_N$ would change much less for the hard direction compared with the easy direction. Moreover, neither the specific heat nor the neutron scattering spectra \cite{Tom87} show evidence of an antiferromagnetic gap over the experimental range, i.\,e., if there is a gap it is well below 1\,K and much smaller than our field range, thus underscoring the isotropic nature of the magnetic state.

The $T^3$ dependence of $C(T)$ may not be accounted for by lattice contributions, which are negligibly small up to 30\,K as estimated from the Debye temperature $\Theta_D = 454$\,K, where $\beta_D \approx 2.1 \cdot 10^{-5}$\,J/molK$^4$ \cite{Bar84}. More surprisingly, an estimated contribution of local-moment antiferromagnetic spin waves to $C(T)$ given by $\Delta C_{sw}(T)=\beta_{sw} T^3$ ($\beta_{sw} \approx 1.65 \cdot 10^{-5}$\,J/mol K$^4$) is two orders of magnitude smaller than experiment, where a conservative estimate of an average $J \approx 1.9$\,meV, derived from the experimentally measured low-$T$ spin-wave velocity $c_{l}$ \cite{Tom87,deJ74}, corresponds to a mean field ordering temperature $T_N \approx 30$\,K, in broad quantitative agreement with experiment. This highlights the presence of strong dynamics in the metallic state challenging conventional models. 

The total entropy $S(T)$ (inset of Fig. 2) of the normal metallic state above $T_N$ and the magnetic order are computed from the experimental data of $C(T)$. Near $T_N$ it reaches {$R$\,ln\,2} per Mn$_3$Si formula unit, i\,e., $S$ is at most one third of the entropy expected of three localised spin 1/2, where the lattice contribution is less than 5\,$\%$. On the other hand, $S(T)$ exceeds that of typical weakly ferromagnetic transition metal compounds by at least an order of magnitude \cite{Mor85,Vis74}. This property fundamentally questions the consistency of the metallic state with the standard spin fluctuation theory in which only paramagnons essentially account for the entropy \cite{Mor85}.

We finally turn to the resistivity $\rho(T)$ (Fig. 3), which displays a broad shoulder near 50\,K \cite{Kal74}. The maximum in $d\rho/dT$ near $T_N$ (inset of Fig. 3) corresponding to the shoulder in $\rho(T)$ shows that it is of electronic origin and closely related to the antiferromagnetic spin fluctuations seen in neutron-scattering well above $T_N$. For $T < 10$\,K, $\rho(T) = \rho_0 + A T^2$ ($\rho_0 \approx 11\mu \Omega$cm, $A \approx 0.115 \mu \Omega$cm/K$^2$) enters a quadratic dependence, where the residual resistivity $\rho_0$ is low, corresponding to a low level of defects. The Kadowaki-Woods ratio $A/ \gamma^2$ is empirically consistent with other materials \cite{Kad86}. As for $C(T)$ and $\chi(T)$, the features of $\rho(T)$ are not sensitive to magnetic field up to 12\,T, even in the immediate vicinity of $T_N$ where the magnetic field energy is very large versus the energy gain of the magnetic order.

\begin{figure}
\centerline{\psfig{file=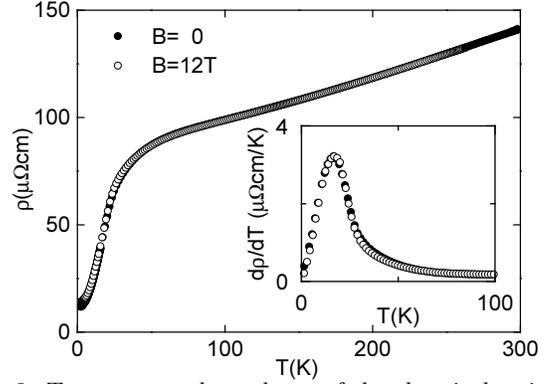,clip=,width=7.5cm}}
\caption{Temperature dependence of the electrical resistivity $\rho(T)$ at zero
magnetic field and $B=12$\,T. The broad shoulder around 50\,K indicates the loss of a dominant scattering mechanism associated with the antiferromagnetic ordering temperature $T_N\approx 23$\,K. The inset displays the derivative of the resisitivity $d \rho /dT$, where the maximum falls just below $T_N$.}
\end{figure}

Although numerous itinerant electron magnets display large ordered moments, e. g. SrRuO$_3$ \cite{sin}, one needs to consider the possibility that Mn$_3$Si has in fact localized moments for the Mn$_I$ sublattice. However, our estimate of $J \approx 1.9$\,meV predicts a large response of the Mn$_I$ moments in a magnetic field, contrary to what is observed. These considerations and the specific heat clearly rule out a Heisenberg model as being even remotely suitable to explain the properties observed here.

The strong interplay of the resistivity and the magnetism and the small change of entropy at $T_N$ as well as the inelastic neutron scattering spectra 
\cite{Tom87} unambiguously show that the magnetic order in Mn$_3$Si is indeed intimately connected with the metallic state. Moreover, electronic structure calculations indicate that the d-electrons form the conduction electron bands \cite{Fuj95,Vla90,Moh98}. However, the magnetisation of Mn$_3$Si is exceptional because it is {\it linear} over the entire field and temperature range studied. 
This shows that there is no mode-mode coupling between ${\bf q}=0$ and $\bf Q$
despite the low ordering temperature \cite{mode}. The invariance of $C(T)$ and $\rho(T)$ with $B$ supports the latter point. Yet, the absence of mode-mode coupling in an itinerant antiferromagnet drastically contrasts the basic assumptions of a spin polarized Fermi liquid, namely, that the mode-mode coupling is $\bf q$-independent \cite{Mor85}.

Measurements of $\rho(T)$ as function of pressure show that $T_N$ increases as $dT_N/dp \approx 0.4$\,K/kbar, in agreement with the Ehrenfest relation \cite{Kal74,Bar84} and many other Heusler compounds \cite{Shi87}. This contrasts the majority of weakly magnetic transition metal, actinide and lanthanide compounds \cite{Pro96}. The $p$-dependence suggests a possible link with two further unsettled condensed matter problems: (i) the magnetic and metallic state related to the quantum phase transition on the "left-hand-side" of the Doniach phase diagram, i.e. for low exchange coupling $J$ and nearly localised moments \cite{Don74}, and (ii) the metallic state on the underdoped side of the superconductivity dome of the cuprates \cite{DDWA}. Our study suggests that these regimes are not compatible with Fermi liquid theory. 

For the remainder we consider the possible origin of deviations from Fermi liquid behaviour in Mn$_3$Si. To date special features of the Fermi surface topology have not been considered in a phenomenological, quantitative evaluation of this model \cite{Mor85}. But electronic structure calculations show that the Fermi surface of Mn$_3$Si is exceptional in two ways: (i) it is strongly nested so that the magnetic order may be understood as highly anharmonic spin density wave (SDW) \cite{Vla90,Moh98}, and (ii) in the spin polarized state the minority density of states of the Mn$_I$ and the Mn$_{II}$ 
are strongly gapped and nearly gapped, respectively, as for half-metallic magnetism \cite{Fuj95}. Consideration of these features suggest fundamental differences of metallic magnetism in Mn$_3$Si with conventional theory.

For a SDW state with a nested, multisheeted Fermi surface an abundance of 
interband excitations is expected. This would be consistent with many features of the unusual inelastic neutron scattering spectra of Mn$_3$Si as noted before \cite{Tom87}, including a damping mechanism other than the particle-hole continuum and spin wave interactions, e.g. a $\bf q$-dependent mechanism in the vicinity of ${\bf Q}$ \cite{Tom87,Yam95}. First principles calculations of excitation spectra in metals support this point in general, as they are extremely sensitive to details of the band structure giving way to interband transitions \cite{Coo79}. Further, it is possible to show that the imaginary part of the experimentally measured dynamical susceptibility of the low energy modes corresponds to an equation of motion that describes retarded, nonlocal diffusion \cite{Boe1}. This behavior is far removed from a propagating spin wave and classical spin wave models are clearly not applicable.

It has long been pointed out, that the antiferromagnetic order and the 
excitation spectra of Mn$_3$Si share many features with Cr \cite{Tom87}, 
where a SDW competes with a charge-density wave and lattice instability \cite{Faw88,Burk}. The amplitude variation of the ordered moment in Mn$_3$Si by one order of magnitude may be a strong hint at the vicintiy to a charge density wave instability. We therefore speculate if an abundance of charge density fluctuations suppresses $T_N$. The metallic state may then be ultimately better described from the point of view of incipient spin-charge separation common to low-dimensional systems \cite{Sch99}, which however in opur case would occur in a 3-dimensional state. In this respect our experiments may also provide an important clue to the long-standing issue of magnetic ordering in Cr, where $T_N$ is so large that the complete lack of field dependence does not seem a surprise for the highest field strengths 
of order 20\,T studied to date \cite{Bar81}.

The second important topological feature of the electronic structure of Mn$_3$Si is, that the minority bands in the spin polarized state are gapped at the Fermi level \cite{Fuj95}. Thus low energy particle-hole excitations do not involve a spin flip and are hence insensitive to magnetic field. Quasiparticle excitations here carry charge but no spin alongside non-QP excitations \cite{Pic01}. This state is referred to as half-metallic magnetism, for which the absence of a magnetic field dependence, as observed here, is key evidence. Besides the electronic structure calculations for Mn$_3$Si the class of Heusler and semi-Heusler systems such as Fe$_2$MnSi \cite{Fuj95} have long been suspected to supply prime candidates for half-metallic magnetism \cite{Pic01}. However, Mn$_3$Si represents the first experimental candidate of an antiferromagnetic
half-metal \cite{Leu95}.

The highly unusual combination of magnetic, structural and metallic properties 
of Mn$_3$Si allow to conclude, that the microscopic underpinning of the overall spectrum of low lying excitations is radically different from that conventionally assumed in Fermi liquid theory or the Heisenberg model. This calls for alternative theoretical input possibly related to incipient spin-charge separation or half metallic magnetism.

We would like to thank G. G. Lonzarich, M. B. Maple, A. Rosch, A. Schr\"oder, 
O. Stockert, M. Uhlarz, P. W\"olfle and V. Ziebat. J. B. would like to acknowledge financial support by the Deutsche Akademische Austauschdienst.
Financial support by the Deutsche Forschungsgemeinschaft is also achnowledged.


\vspace{-5mm}
\begin{references}
\vspace{-17mm}

\bibitem{Mor85} T. Moriya, {\it Spin fluctuations in itinerant electron magnetism} Springer-Verlag, Berlin (1985); G. G. Lonzarich, L. Taillefer, J. Phys. C {\bf 18} 4339 (1985).

\bibitem{Alk95} H. J. Al-Kanani, J. G. Booth, J. Mag. Mag. Mat. {\bf 140-144} (1995) 1539.

\bibitem{Str77} E. Stryjewski, N. Giordano, Adv. Phys. {\bf 26} (1977) 487.

\bibitem{MMT} {\it Handbook of the Physics and Chemistry of Rare Earths}, ed.: K. A. Gschneidner jr. and L. Eyring, Elsevier, Amsterdam. 

\bibitem{Kub83} J. K\"ubler et al., Phys. Rev. B {\bf 28} (1983) 1745; P. J. Webster, Contemp. Phys. {\bf 10} (1969) 559.

\bibitem{Fuj95} S. Fujii et al., J. Phys. Soc. Jpn. {\bf 64} (1995) 185.

\bibitem{Nis97} Y. Nishino et al., Phys. Rev. Lett. {\bf 79} (1997) 1909; Ye Feng et al., Phys. Rev. B {\bf 63} (2001) 165109.

\bibitem{Sle00} A. Slebarski et al., Phys. Rev. B {\bf 62} (2000) 3296.

\bibitem{Kal74} G. I. Kalishevich et al., Sov. Phys. Sol. St. {\bf 16} (1974) 1151.

\bibitem{Bab75} E. N. Babanova et al., Sov. Phys. Sol. St. {\bf 17} (1975) 616.

\bibitem{Tom75} S. Tomiyoshi et al., J. Phys. Soc. Jpn. {\bf 39} (1975) 295.

\bibitem{Tom87} S. Tomiyoshi et al., Phys. Rev. B {\bf 36} (1987) 2181.

\bibitem{Yam95} Y. Yamaguchi et al., Physica B {\bf 213 \& 214} (1995) 363.

\bibitem{Col89} M. F. Collins, {\it Magnetic critical scattering}, 
Oxford University Press, Oxford (1989)

\bibitem{Bar84} S. M. Barmin, A. A. Sevast'yanov, Sov. Phys. Sol. St. {\bf 26} (1984) 917.

\bibitem{deJ74} L.J. de Jongh, A. R. Miedema {\it Exp. on Simple
Mag. Model Systems}, Taylor and Francis, London (1974).

\bibitem{Vis74} R. Viswanathan, J. Phys. F: Metal Phys. {\bf 4} (1974) L57.

\bibitem{Kad86} K. Kadowaki, S. B. Woods, Sol. St. Comm. {\bf 58} (1986) 507.

\bibitem{sin} D. J. Singh,  J. Appl. Phys. {\bf 79} (1996) 4818.

\bibitem{Vla90} A. Vlasov et al., Sov. Phys. Leb. Inst. Rep. {\bf 12} (1990) 5.

\bibitem{Moh98} P. Mohn, E. Supanetz, Phil. Mag. B {\bf 78} (1998) 629.

\bibitem{mode}
Nonlinearities of $M(B)$ derive from higher order contributions to the free energy, which imply in the unconditional probability distribution of the partition function the coupling of magnetisation modes ${\bf m}({\bf q},\omega)$ \cite{Mor85}.

\bibitem{Pro96} See e. g. J. Phys.: Cond. Matter {\bf 8} (1996) 9675.
 
\bibitem{Shi87} K. Shirakawa et al., J. Mag. Mag. Mat. {\bf 70} (1987) 421.

\bibitem{Don74} S. Doniach, Physica B {\bf 91} (1977) 213.

\bibitem{DDWA} W. Liang, J. Phys.: Condens. Matter {\bf 10} (1998) 11365.

\bibitem{Coo79} J. F. Cooke, J. Appl. Phys.  {\bf 50}  (1979) 7439.

\bibitem{Boe1} J. B{\oe}uf, C. Pfleiderer, unpublished (2001); D. Forster {Hydrodynamic Fluctuations, Broken Symmetry and Correlation Functions}, W. A. Benjamin, Reading MA (1975).

\bibitem{Faw88} E. Fawcett, E., Rev. Mod. Phys. {\bf 60} (1988) 209.

\bibitem{Burk} C. R. Fincher jr. et al., Phys. Rev B {\bf 24} 1312 (1981).

\bibitem{Sch99} A. Schofield, Contemp. Physics {\bf 40} (1999) 95. 

\bibitem{Bar81} Z. Barak et al., J. Phys.: Metal Phys. {\bf 11} (1981) 915.

\bibitem{Pic01} V. Y. Irkhin, M. I. Katsnel'son, Pys.-Usp. {\bf 37} (1994) 659.

\bibitem{Leu95} H. van Leuken, R. A. de Groot, Phys. Rev. Lett. {\bf 74} (1995) 1171.

\end{references}
\end{document}